\documentclass[11pt]{article}

\usepackage[nosort]{cite}

\usepackage[british]{babel} 

\usepackage{epsfig,rotating}
\usepackage{amsmath,amssymb,amsthm}
\usepackage{amsfonts}
\usepackage{mathrsfs}
\usepackage{bbm}
\usepackage{bm}

\usepackage{color}
\usepackage[textwidth = 430 pt, textheight = 630 pt]{geometry}
\definecolor{MyDarkBlue}{rgb}{0.15,0.25,0.45}

\usepackage[linktocpage=true]{hyperref}
\hypersetup{
colorlinks=true,
citecolor=MyDarkBlue,
linkcolor=MyDarkBlue,
urlcolor=MyDarkBlue,
pdfauthor={TBI},
pdftitle={TBI},
pdfsubject={hep-th}
breaklinks=true
}

\let\fn\footnote
\renewcommand{\footnote}[1]{\linespread{1.1}\fn{#1}\linespread{1.29}}

\makeatletter\renewcommand{\section}{\@startsection
{section}{1}{\z@}{-3.5ex plus -1ex minus
    -.2ex}{2.3ex plus .2ex}{\bf\mathversion{bold} }}
\makeatletter\renewcommand{\subsection}{\@startsection{subsection}{2}{\z@}{-3.25ex
plus -1ex minus
   -.2ex}{1.5ex plus .2ex}{\bf\mathversion{bold} }}
\makeatletter\renewcommand{\subsubsection}{\@startsection{subsubsection}{3}{\z@}{-3.25ex
plus -1ex minus -.2ex}{1.5ex plus .2ex}{\it }}
\renewcommand{\thesection}{\arabic{section}}
\renewcommand{\thesubsection}{\arabic{section}.\arabic{subsection}}
\renewcommand{\@seccntformat}[1]{\@nameuse{the#1}.~~}

\renewcommand{\theequation}{\thesection.\arabic{equation}}
\makeatletter \@addtoreset{equation}{section}

\renewcommand*\l@section{\@dottedtocline{1}{0em}{2em}}
\renewcommand*\l@subsection{\@dottedtocline{2}{2em}{2.4em}}
\renewcommand*\l@subsubsection{\@dottedtocline{4}{3.8em}{3.7em}}

\renewcommand\tableofcontents{%
    \section*{\large\contentsname
        \@mkboth{%
          \MakeUppercase\contentsname}{\MakeUppercase\contentsname}}%
       {\baselineskip=15pt plus 2pt minus 1pt
   \@starttoc{toc}}%

}

\newcommand{\acknowledgements}{\section*{Acknowledgements}
\addcontentsline{toc}{section}{\hspace{0.6cm}{\bf Acknowledgements}}}

\newcommand{\appendices}{
\section*{Appendix}\label{appendices}\setcounter{subsection}{0}
\addcontentsline{toc}{section}{Appendix}
\setcounter{equation}{0}
\makeatletter
\renewcommand{\theequation}{\Alph{subsection}.\arabic{equation}}
\renewcommand{\thesubsection}{\Alph{subsection}}
\@addtoreset{equation}{subsection}
\makeatother
}

\theoremstyle{plain}
\newtheorem{thm}{Theorem}[section]

\newtheorem{prop}[thm]{Proposition}
\newtheorem{exm}[thm]{Example}
\newtheorem{defn}[thm]{Definition}
\newtheorem{rem}[thm]{Remark}

\def\periodb#1{\setbox0=\hbox{$#1$}#1\hskip-\wd0\hbox to\wd0{-}}

\def\tyng(#1){\hbox{\tiny$\yng(#1)$}}           
\def\tyoung(#1){\hbox{\tiny$\young(#1)$}}           

\begin{document}
\begin{titlepage}

\setcounter{page}{0}
\renewcommand{\thefootnote}{\fnsymbol{footnote}}

\begin{flushright}
\end{flushright}

\vspace{1cm}

\begin{center}

{\LARGE\textbf{\mathversion{bold}Relations between symmetries and
conservation laws for difference systems}\par}

\vspace{0.5cm}

{\large Linyu Peng\footnote{{\it E-mail address:\/} \href{mailto
l.peng@aoni.waseda.jp}{\ttfamily l.peng@aoni.waseda.jp}}}

\vspace{0.5cm}
%

%
 {
Department of Mathematics, University of Surrey, Guildford GU2 7XH,
United Kingdom\\
{\it Current address}: Department of Applied Mechanics and Aerospace
Engineering, Waseda University, Ohkubo, Shinjuku, Tokyo 169-8555,
Japan}

\vspace{0.5cm}

{\bf Abstract}
\end{center}
\vspace{-.3cm}

For difference variational problems on lattice, this paper presents
a relation between divergence variational symmetries and
conservation laws for the associated Euler-Lagrange system provided
by Noether's theorem. This hence inspires us to define conservation
laws related with symmetries for arbitrary difference equations with
or without Lagrangian formulations. These conservation laws are
constrained by some partial differential equations obtained from the
symmetries generators. It is shown that the orders of these partial
differential equations have been reduced comparing to the ones used
in a general approach. Illustrating examples are presented.
\begin{quote}

 \vfill

\noindent\today

\end{quote}

\setcounter{footnote}{0}\renewcommand{\thefootnote}{\arabic{thefootnote}}

{\bf Keywords:} difference variational problem; difference equation;
symmetry; conservation law

\end{titlepage}

 \tableofcontents

\bigskip
\bigskip
\hrule
\bigskip
\bigskip

\section{Introduction}
For differential variational problems, their variational symmetries
are still symmetries of the associated Euler-Lagrange system and the
converse is not necessary to be true \cite{Olver:1993aa}.
Nevertheless, Dorodnitsyn
\cite{Dorodnitsyn:2001aa,Dorodnitsyn:2010aa} proved that the
invariance of a difference Lagrangian from the mesh viewpoint does
not mean the invariance of the corresponding difference
Euler-Lagrange equations. The first task of this paper is to show
that divergence variational symmetries of difference variational
problems from the lattice viewpoint are still symmetries of the
underlying difference Euler-Lagrange equations.

For differential systems, the well-known Noether's first theorem
\cite{Noether:1918aa} tells us that variational symmetries will
amount to conservation laws for differential variational problems.
The relationship between the variational symmetries and the
associated conservation laws can be explicitly determined
\cite{Olver:1993aa,Goncalves:2012aa,Ibragimov:1998aa}. If a
differential system does not have a underlying Lagrangian
formulation, to what extend can we relate its symmetries and
conservation laws? To answer this question, for self-adjoint
differential equations, Anco and Bluman \cite{Anco:1996aa} derived
an equality providing a correspondence between symmetries and
conservation laws and it does not depend on a Lagrangian
formulation. In \cite{Kara:2000aa}, Kara and Mahomed presented a
method showing a relationship between conservation laws and symmetry
generators for an arbitrary differential equation. Moreover,
conservation laws of differential equations may also be obtained by
use of operators which are not symmetry generators of the associated
system \cite{Kara:2006aa}.

The other part of this paper hence contributes to find out a
relation between symmetries for difference variational problems and
their Noether-type of conservation laws. To some extend, this can be
generalized to difference equations without Lagrangian formulations.
Given a group of Lie point symmetries, usually one can easily
calculate the related conservation laws by solving partial
differential equations. The order of these partial differential
equations is reduced comparing to the method provided by Rasin \&
Hydon \cite{Rasin:2007aa}. This method is illustrated by some
ordinary difference equations as well as partial difference
equations, such as equations of the ABS classification
\cite{Adler:2003bs}.

\section{Symmetries and conservation laws for difference systems}
Let $n=(n^1,n^2,\ldots,n^p)\in \mathbb{Z}^p$ and
$u_n=(u_n^1,u_n^2,\ldots,u_n^q)\in U\subset \mathbb{R}^q$ be the
independent and dependent variables respectively. The shift operator
(or map) $S$ is defined as
\begin{equation}
\begin{aligned}
S_k:n^i\mapsto n^i+\delta_k^i,~~k=1,2,\ldots,p,
\end{aligned}
\end{equation}
where $\delta^i_k$ is the Kronecker delta. Let $\bold{1}_k$ be the
$p$-tuple with only one nonzero entry $1$ in the $k$-th place. Then
the $k^{\operatorname{th}}$-shift map and the naturally extended
action to a function $f(n)$ are respectively given by
\begin{equation}
\begin{aligned}
S_k:n\mapsto n+\bold{1}_k,
\end{aligned}
\end{equation}
and
\begin{equation}
\begin{aligned}
S_k:f(n)\mapsto f(n+\bold{1}_k).
\end{aligned}
\end{equation}
We will use the notation $S_{\bold{1}_k}$ instead of $S_k$, and the
composite of shifts using multi-index notation is given by $S_J$,
where $J$ is a $p$-tuple. We can define the inverse of the shift map
as $S_{-\bold{1}_k}:n\to n-\bold{1}_k$. The inverse map $S_{-J}$ of
the composite of shifts is similarly defined.

Consider the following one-parameter transformations (Lie point
transformations)
\begin{equation}
\Gamma_{\varepsilon}:(n,u_n)\mapsto(n,\tilde{u}_n),
\end{equation}
with $\tilde{u}_n=\tilde{u}(n,u_n;\varepsilon)$. In the present
paper, the independent variable is always invariant, that is,
$\tilde{n}=n$. This admits a Taylor expansion with respect to
$\varepsilon$,
\begin{equation}\label{tay0}
\tilde{u}^{\alpha}_n=u^{\alpha}_n+\varepsilon
Q^{\alpha}(n,u_n)+O(\varepsilon^2).
\end{equation}
From this, the infinitesimal generator is obtained as
\begin{equation}
\bold{v}=Q^{\alpha}_n\frac{\partial}{\partial u_n^{\alpha}},
\end{equation}
where $Q_n=(Q_n^1,Q_n^2,\ldots,Q^q_n)$ is called a characteristic,
and $Q^{\alpha}_n=Q^{\alpha}(n,u_n)$. The Einstein summation
convention is applied here and all through this paper. For higher
transformations, $Q_n$ is a $q$-tuple defined on $n,u_n$ as well as
their shifts. The transformation $\Gamma_{\varepsilon}$ can be
prolonged to shifts of $u_n$; we have the (infinitely) prolongation
of the vector field $\bold{v}$ as
\begin{equation}
\bold{pr}^{(\infty)}\bold{v}=\sum_{|J|\geq0}^{\infty}Q^{\alpha}_{n+J}\frac{\partial}{\partial
u_{n+J}^{\alpha}},
\end{equation}
with $Q^{\alpha}_{n+J}:=S_JQ_n^{\alpha}$. For the sake of
simplicity, we also write $Q^{\alpha}_{n+J}$ as $Q^{\alpha}_J$,
$u^{\alpha}_{n+J}$ as $u^{\alpha}_J$.

Consider a system of difference equations
\begin{equation}
\mathcal{A}^{\vartriangle}=\{F_k=0,~~k=1,2,\ldots,l\}, \label{fde}
\end{equation}
where $F_k$ are functions of $n$, $u_n^{\alpha}$ and their shifts. A
one-parameter group $G$ of transformations $\Gamma_{\varepsilon}$
related to the system is a symmetry group, if all the associated
infinitesimal generators $\bold{v}$ satisfy
\begin{equation}
\bold{pr}^{(\infty)}\bold{v}(F_k)=0,~~\text{when (\ref{fde}) holds}.
\end{equation}

Hydon \cite{Hydon:2012aa} provided a systematic method (by using the
linearized symmetry condition) for determining one-parameter Lie
groups of symmetries for ordinary difference equations.

Consider a $k^{\operatorname{th}}$-order ordinary difference
equation given by
\begin{equation}\label{kdode}
u_{n+k}=\omega(n,u_n,u_{n+1},\ldots,u_{n+k-1}),~~\text{with}~~\frac{\partial\omega}{\partial
u_n}\neq0.
\end{equation}
By writing the infinitesimal generator as
\begin{equation}
\bold{v}=\sum_m\left(S^mQ_n\right)\frac{\partial}{\partial u_{n+m}},
\end{equation}
where the characteristic $Q_n$ is a function of
$n,u_n,u_{n+1},\ldots,u_{n+k-1}$, the linearized symmetry condition
gives rise to
\begin{equation}\label{cfelsc}
S^kQ_n=\bold{v}(\omega).
\end{equation}
This is the necessary and sufficient condition to obtain (local)
one-parameter Lie groups of symmetries generated by $\bold{v}$.

 For a system of
difference equations given by (\ref{fde}), a {\it conservation law}
is defined as a difference divergence expression
\begin{equation}
\operatorname{Div}^{\vartriangle}P:=\sum_{i=1}^p(S_i-\operatorname{id})P^i,
\end{equation}
which vanishes on all solutions of the system, where $P$ is a
$p$-tuple defined on $n,u_n$ as well as the shifts of $u_n$, and for
any $i$, $P^i$ is smooth with respect to continuous variables.

A conservation law can be trivial in two ways: the $p$-tuple $P$
vanishes on solutions, that is, $P|_{\mathcal{A}^{\vartriangle}}=0$,
or $\operatorname{Div}^{\vartriangle}P\equiv0$, without reference to the system. A
conservation law is trivial if it is a linear superposition of these
two types of trivial conservation laws.

For ordinary difference equations, i.e., $p=1$, a conservation law
is also called a first integral. Consider the
$k^{\operatorname{th}}$-order difference equation (\ref{kdode}); a
first integral is a nonconstant function
\begin{equation}
P=P(n,u_n,u_{n+1},\ldots,u_{n+k-1})
\end{equation}
 satisfying
\begin{equation}\label{dcffi}
SP=P,~~\text{when}~~(\ref{kdode})~~\text{holds},
\end{equation}
which can be considered as a determining condition for first
integrals.

Let $G$ be a one-parameter Lie group acting on $U$ as point
transformations; write this action as $\tilde{u}_n=g\circ u_n$ for
$g\in G$ . For a difference variational problem
\begin{equation}\label{dl1}
\mathscr{L}[u]=\sum_nL(n,[u])=\sum_nL(n,u_n^{\alpha},u_{n+\bold{1}_i}^{\alpha},\ldots,u_{n+J}^{\alpha}),
\end{equation}
its Lagrangian is said to be invariant under $G$ if
\begin{equation}
L(n,\tilde{u}_n^{\alpha},\tilde{u}_{n+\bold{1}_i}^{\alpha},\ldots,\tilde{u}_{n+J}^{\alpha})=L(n,u_n^{\alpha},u_{n+\bold{1}_i}^{\alpha},\ldots,u_{n+J}^{\alpha})
(=L_n).
\end{equation}
Here $[u]$ denotes $u_n$ and its shifts. This has an infinitesimal
version
\begin{equation}
\bold{pr}^{(\infty)}\bold{v}(L_n)=0,
\end{equation}
with $\bold{v}=Q^{\alpha}_n\frac{\partial}{\partial u_n^{\alpha}}$
the infinitesimal generator. It can be generalized to {\it
divergence variational symmetries} satisfying
\begin{equation}
\bold{pr}^{(\infty)}\bold{v}(L_n)=\sum_{i}(S_i-\operatorname{id})R^i,
\end{equation}
which again keeps $\mathscr{L}[u]$ invariant. A discrete version of
Noether's theorem exists (see e.g.,
\cite{Grant:2011aa,Grant:2013hy}); it relates symmetries for
difference variational problems and conservation laws of the
associated Euler-Lagrange equations.

The following proposition is useful in the proofs for the coming
results.

 \begin{prop}
For any function $f(n,[u])$, we have
\begin{enumerate}
\item $\frac{\partial}{\partial u^{\alpha}_{J_0}}(S_{-J}f)=S_{-J}\left(\frac{\partial f}{\partial
u^{\alpha}_{J_0+J}}\right)$, for $\forall \alpha$ and $J_0,J$;
\item $S_i\left(\bold{pr}^{(\infty)}\bold{v}(f)\right)=\bold{pr}^{(\infty)}\bold{v}\left(S_i f\right)$.
\end{enumerate}
 \end{prop}

 \begin{rem}
The proposition above implies that the divergence operator
$\operatorname{Div}^{\vartriangle}$ and the symmetry generator are commutative,
\begin{equation}
\operatorname{Div}^{\vartriangle}\left(\bold{pr}^{(\infty)}\bold{v}(P)\right)=\bold{pr}^{(\infty)}\bold{v}\left(\operatorname{Div}^{\vartriangle}P\right).
\end{equation}
Here $P$ is a $p$-tuple.
 \end{rem}


 The difference
Euler-Lagrange operator is given by
\begin{equation}
\bold{E}_{\alpha}^{\vartriangle}:=\sum_JS_{-J}\frac{\partial}{\partial
u_{J}^{\alpha}},
\end{equation}
and the Euler-Lagrange equations with respect to the variational
problem (\ref{dl1}) can be simply written as
\begin{equation}
\bold{E}_{\alpha}^{\vartriangle}(L_n)=0.
\end{equation}

For a Lagrangian $L_n$, let us define
$L^{(\infty)}=\sum_JS_{-J}L_n$; then obviously one can check that
\begin{equation}
\bold{E}_{\alpha}^{\vartriangle}(L_n)=\frac{\partial}{\partial
u^{\alpha}}L^{(\infty)}.
\end{equation}

\begin{rem}\label{remlc}
\begin{enumerate}
\item If $u_n\mapsto v_n$ is Lie point transformations, then
\begin{equation}
\bold{E}_{u^{\alpha}}^{\vartriangle}=\frac{\partial
v^{\beta}}{\partial u^{\alpha}}\bold{E}_{v^{\beta}}^{\vartriangle}.
\end{equation}
\item If $\bold{v}=Q^{\alpha}\partial_{u^{\alpha}}$ generates a group of divergence variational
symmetries of $\sum_nL_n$, then
\begin{equation}
\bold{v}\left(L^{(\infty)}\right)=Q^{\alpha}\bold{E}_{u^{\alpha}}^{\vartriangle}(L_n)(=\operatorname{Div}^{\vartriangle}P_0),
\end{equation}
for some $p$-tuple $P_0$.
\end{enumerate}
\end{rem}

Integrating by parts gives that
\begin{equation}\label{ee1}
\begin{aligned}
\bold{pr}^{(\infty)}\bold{v}(L_n)&=\sum_{\alpha,
J}Q^{\alpha}_{J}\frac{\partial
L_n}{\partial u_{J}^{\alpha}}\\
&=\sum_{\alpha, J}Q^{\alpha}S_{-J}\left(\frac{\partial L_n}{\partial
u_{J}^{\alpha}}\right)+\sum_i(S_i-\operatorname{id})C^i\\
&=Q^{\alpha}\bold{E}_{\alpha}^{\vartriangle}(L_n)+\sum_i(S_i-\operatorname{id})C^i,
\end{aligned}
\end{equation}
where Noether's conservation law is given by components as
\begin{equation}\label{ee2}
P^i=C^i-R^i=\sum_{\alpha}\sum_{J\geq
\bold{1}_i}S_{-\bold{1}_i}\left(\bold{E}^{\vartriangle}_{u_J^{\alpha}}(L_n)\right)Q^{\alpha}_{J-\bold{1}_i}-R^i;
\end{equation}
here $\bold{0}$ is a $p$-tuple with all components zeros, and
the operator $\bold{E}^{\vartriangle}_{u_J^{\alpha}}$ is given by
\begin{equation}
\bold{E}^{\vartriangle}_{u_J^{\alpha}}(L_n)=\sum_{J_0\geq\bold{0}}S_{-J_0}\left(\frac{\partial
L_n}{\partial u_{J_0+J}^{\alpha}}\right).
\end{equation}
The formula (\ref{ee2}) is only valid for
Lagrangians which are independent from backward shifts of $u_n$.
Nevertheless, this can be always achieved by shifting Lagrangians
forward, without affecting solutions of the associated
Euler-Lagrange equations \cite{Peng:2012aa,Peng:2013ad}. In this paper, we will only consider Lagrangians which are independent from backward shifts of $u$. Nevertheless, we have the following remark for another kind of Lagrangians.

\begin{rem}
For Lagrangians which depende on $u_n$ and its backward shifts only, we
will have the conservation law as
\begin{equation}
C^i=-\sum_{\alpha}\sum_{J<\bold{0}}\widetilde{\bold{E}}^{\vartriangle}_{u_J^{\alpha}}(L_n)Q^{\alpha}_J.
\end{equation}
However, the operator $\widetilde{\bold{E}}^{\vartriangle}_{u_J^{\alpha}}$ is
\begin{equation}
\widetilde{\bold{E}}^{\vartriangle}_{u_J^{\alpha}}(L_n)=\sum_{J_0\leq\bold{0}}S_{-J_0}\left(\frac{\partial
L_n}{\partial u_{J_0+J}^{\alpha}}\right).
\end{equation}
\end{rem}

\section{Relations between symmetries and conservation laws}
 To find out how divergence variational symmetries affect the
Euler-Lagrange equations, we recall that
$\sum_i(S_i-\operatorname{id})P^i_0$ is a null Lagrangian such that
$\bold{E}_{\alpha}^{\vartriangle}\left(\sum_i(S_i-\operatorname{id})P^i_0\right)$
vanishes always. Then we get the following identity
\begin{equation}\label{ninini}
\begin{aligned}
\bold{E}_{\alpha}^{\vartriangle}\left(\bold{pr}^{(\infty)}\bold{v}(L_n)\right)&
=\bold{E}_{\alpha}^{\vartriangle}\left(Q^{\beta}\bold{E}_{\beta}^{\vartriangle}(L_n)\right)\\
&=\sum_JQ^{\beta}_{-J}S_{-J}\left(\frac{\partial
\bold{E}_{\beta}^{\vartriangle}(L_n)}{\partial
u_{J}^{\alpha}}\right)+\sum_JS_{-J}\left(\frac{\partial
Q^{\beta}}{\partial
u_{J}^{\alpha}}\bold{E}_{\beta}^{\vartriangle}(L_n)\right).
\end{aligned}
\end{equation}
The first term on the right-hand side of (\ref{ninini}) can be
rearranged as 
\begin{equation}
\begin{aligned}
\sum_JQ^{\beta}_{-J}S_{-J}\left(\frac{\partial
\bold{E}_{\beta}^{\vartriangle}(L_n)}{\partial
u_{J}^{\alpha}}\right)&=\sum_{J_0,J}Q^{\beta}_{-J}S_{-J}\left[\frac{\partial}{\partial
u_{J}^{\alpha}}\left(S_{-J_0}\left(\frac{\partial L_n}{\partial u_{J_0}^{\beta}}\right)\right)\right]\\
&=\sum_{J_0,J}Q^{\beta}_{-J}S_{-J-J_0}\left(\frac{\partial^2L_n}{\partial
u_{J_0}^{\beta}u_{J+J_0}^{\alpha}}\right)\\
&=\sum_{J_0,J}Q^{\beta}_{-J}\frac{\partial}{\partial
u_{-J}^{\beta}}\left(S_{-J-J_0}\left(\frac{\partial L_n}{\partial
u_{J+J_0}^{\alpha}}\right)\right)\\
&=\sum_{J_1,J}Q^{\beta}_{-J}\frac{\partial}{\partial
u_{-J}^{\beta}}\left(S_{-J_1}\left(\frac{\partial L_n}{\partial
u_{J_1}^{\alpha}}\right)\right)\\
&=\bold{pr}^{(\infty)}\bold{v}(\bold{E}_{\alpha}^{\vartriangle}(L_n)).
\end{aligned}
\end{equation}
Here we set $J_1=J+J_0$ on the fourth equality. Therefore, we have
proved the following theorem for a one-parameter group of divergence
variational symmetries as the second term from the right-hand side
of (\ref{ninini}) only depends on shifts of the Euler-Lagrange
equations. Nevertheless, this proof procedure can be generalized to
any finite-dimensional groups of divergence variational symmetries.

\begin{thm}
Any group of divergence variational symmetries of a Lagrangian on
lattice $\mathbb{Z}^p$ is still a group of symmetries for the
associated difference Euler-Lagrange equations.
\end{thm}

Rewrite Noether's conservation law as follows by using operators
$T^i$
\begin{equation}
P^i=C^i-R^i=T^i(L_n)-R^i,
\end{equation}
where
\begin{equation}
T^i=\sum_{\alpha}\sum_{J\geq\bold{1}_i}Q^{\alpha}_{J-\bold{1}_i}S_{-\bold{1}_i}\bold{E}^{\vartriangle}_{u^{\alpha}_J}.
\end{equation}

\begin{thm}
For Lie point variational symmetries, implying $R^i=0$, their
infinitesimal generators satisfy that
\begin{equation}
0=\bold{pr}^{(\infty)}\bold{v}(P^i)\left(=\bold{pr}^{(\infty)}\bold{v}(T^i(L_n))\right),~~\text{for
all}~~i.
\end{equation}
\end{thm}

\begin{proof}
It is sufficient to verify that
\begin{equation}
\bold{pr}^{(\infty)}\bold{v}(T^i(L_n))-T^i\left(\bold{pr}^{(\infty)}\bold{v}(L_n)\right)=0.
\end{equation}
By using Leibniz rule, terms on the left-hand side can be expanded
respectively as
\begin{equation*}
\begin{aligned}
\bold{pr}&^{(\infty)}\bold{v}(T^i(L_n))=\sum_{I,J\geq\bold{1}_i}Q^{\alpha}_I\frac{\partial}{\partial u^{\alpha}_I}\left[Q_{J-\bold{1}_i}^{\beta}
S_{-\bold{1}_i}\bold{E}^{\vartriangle}_{u_J^{\beta}}(L_n)\right]\\
&=\sum_{I,J\geq\bold{1}_i}\left[Q_I^{\alpha}\frac{\partial Q_{J-\bold{1}_i}^{\beta}}{\partial u_I^{\alpha}}S_{-\bold{1}_i}\bold{E}^{\vartriangle}_{u_J^{\beta}}(L_n)+Q_I^{\alpha}Q_{J-\bold{1}_i}^{\beta}S_{-\bold{1}_i}\left(\frac{\partial}{\partial u_{I+\bold{1}_i}^{\alpha}}\bold{E}^{\vartriangle}_{u_J^{\beta}}(L_n)\right)\right]\\
&=\sum_{I,J\geq\bold{1}_i,J_0\geq\bold{0}}\left[Q_I^{\alpha}\frac{\partial Q_{J-\bold{1}_i}^{\beta}}{\partial u_I^{\alpha}}S_{-\bold{1}_i-J_0}\frac{\partial L_n}{\partial u_{J+J_0}^{\beta}}+Q_{J-\bold{1}_i}^{\alpha}Q_I^{\beta}S_{-\bold{1}_i-J_0}\frac{\partial^2L_n}{\partial u_{I+\bold{1}_i+J_0}^{\beta}\partial u_{J+J_0}^{\alpha}}\right]
\end{aligned}
\end{equation*}
(a change of indices has been applied) and
\begin{equation*}
\begin{aligned}
T^i&\left(\bold{pr}^{(\infty)}\bold{v}(L_n)\right)=\sum_{I,J\geq\bold{1}_i}Q_{J-\bold{1}_i}^{\alpha}S_{-\bold{1}_i}\bold{E}^{\vartriangle}_{u_J^{\alpha}}\left(Q_I^{\beta}\frac{\partial L_n}{\partial u_I^{\beta}}\right)\\
&=\sum_{I,J\geq\bold{1}_i,J_0\geq\bold{0}}\left[Q_{J-\bold{1}_i}^{\alpha}S_{-\bold{1}_i-J_0}\left(\frac{\partial Q_I^{\beta}}{\partial u_{J+J_0}^{\alpha}}\frac{\partial L_n}{\partial u_I^{\beta}}\right)+Q_{J-\bold{1}_i}^{\alpha}Q_{I-J_0-\bold{1}_i}^{\beta}S_{-\bold{1}_i-J_0}\frac{\partial^2 L_n}{\partial u_I^{\beta}\partial u_{J+J_0}^{\alpha}}\right]\\
&=\sum_{I,J\geq\bold{1}_i,J_0\geq\bold{0}}\left[Q_{J-\bold{1}_i}^{\alpha}S_{-\bold{1}_i-J_0}\left(\frac{\partial Q_I^{\beta}}{\partial u_{J+J_0}^{\alpha}}\frac{\partial L_n}{\partial u_I^{\beta}}\right)+Q_{J-\bold{1}_i}^{\alpha}Q_I^{\beta}S_{-\bold{1}_i-J_0}\frac{\partial^2 L_n}{\partial u_{I+\bold{1}_i+J_0}^{\beta}\partial u_{J+J_0}^{\alpha}}\right]
\end{aligned}
\end{equation*}
It is obvious that the terms involving second-order derivatives of
$L_n$ are the same. Since
we assumed that $Q^{\alpha}$ are the characteristics of a group of
Lie point variational symmetries, the following identity holds
\begin{equation}
\frac{\partial Q_{I}^{\beta}}{\partial
u_{J}^{\alpha}}=\frac{\partial Q_{J}^{\beta}}{\partial
u_{I}^{\alpha}}=\delta_{I}^J\frac{\partial Q^{\beta}_J}{\partial
u_J^{\alpha}},~~\forall I,J,
\end{equation}
which finishes the proof.
\end{proof}

\begin{rem}
The proof above implies the following commutative relation
\begin{equation}
[\bold{pr}^{(\infty)}\bold{v},T^i]=0,~~\text{for all}~~
i=1,2,\ldots,p.
\end{equation}
Here $[\cdot,\cdot]$ is the Lie bracket, and $\bold{v}$ generates a
group of Lie point (divergence) variational symmetries.
\end{rem}

If $\bold{v}$ is a divergence symmetry generator such that $R^i$ do
not vanish, then we have
\begin{equation}
\bold{pr}^{(\infty)}\bold{v}(P^i)=T^i\left(\sum_j(S_j-\operatorname{id})R^j\right)-\bold{pr}^{(\infty)}\bold{v}(R^i).
\end{equation}

However, for ordinary difference systems, that is, $p=1$, we can get
an even better result.

\begin{thm}
When $p=1$, for groups of Lie point divergence variational
symmetries, their infinitesimal generators are subject to
\begin{equation}
\bold{pr}^{(\infty)}\bold{v}(P)=0.
\end{equation}
Here the first integral is $P=C-R$.
\end{thm}

\begin{proof}
As we already showed that
\begin{equation}
\bold{pr}^{(\infty)}\bold{v}(C)=T\left(\bold{pr}^{(\infty)}\bold{v}(L_n)\right)=T(SR-R),
\end{equation}
the proof finishes if we can verify
\begin{equation}
T(SR-R)-\bold{pr}^{(\infty)}\bold{v}(R)=0.
\end{equation}
This can be done simply by a change of indices as follows:
\begin{equation}
  \begin{aligned}
    T(SR)&=\sum_mQ_{m-1}^{\alpha}S_{-1}\left(\bold{E}^{\vartriangle}_{u^{\alpha}_{n+m}}(SR)\right)\\
&=\sum_{m,k\geq0}Q_{m-1}^{\alpha}S_{-k-1}\frac{\partial (SR)}{\partial u_{n+m+k}^{\alpha}}\\
&=\sum_{m,k\geq0}Q_{m-1}^{\alpha}S_{-k}\frac{\partial R}{\partial u^{\alpha}_{n+m+k-1}}
  \end{aligned}
\end{equation}
and similarly
\begin{equation}
\begin{aligned}
T(R)&=\sum_{m,k\geq0}Q_{m-1}^{\alpha}S_{-k-1}\frac{\partial R}{\partial u_{n+m+k}^{\alpha}}\\
&=\sum_{m,k\geq1}Q_{m-1}^{\alpha}S_{-k}\frac{\partial R}{\partial u^{\alpha}_{n+m+k-1}}.
\end{aligned}
\end{equation}
Therefore, we get that
\begin{equation}
\begin{aligned}
T(SR-R)&=\sum_m\left(\sum_{k\geq0}-\sum_{k\geq1}\right)Q_{m-1}^{\alpha}S_{-k}\frac{\partial R}{\partial u^{\alpha}_{n+m+k-1}}\\
&=\sum_mQ_{m-1}^{\alpha}\frac{\partial R}{\partial u_{n+m}^{\alpha}}\\
&=\bold{pr}^{(\infty)}\bold{v}(R).
\end{aligned}
\end{equation}
\end{proof}


\begin{defn}
For a difference system $\mathcal{A}^{\vartriangle}$, an
infinitesimal generator $\bold{v}$ with respect to a group of Lie
point symmetries is said to be associated with a conservation law
$P$ if the following equality is satisfied
\begin{equation}\label{clss}
\bold{pr}^{(\infty)}\bold{v}(P^i)=0,~~\text{for
all}~~i,~~\text{when}~~\mathcal{A}^{\vartriangle}~~\text{holds}.
\end{equation}
\end{defn}

For a counterpart of similar results for differential systems, we
recommend \cite{Ibragimov:1998aa,Kara:2000aa,Leach:1981aa}. When
$p=1$, that is, for a $k^{\operatorname{th}}$-order ordinary
difference equation (\ref{kdode}), the equality (\ref{clss}) can be
written as the following linear first-order partial differential
equation,
\begin{equation}\label{pde1}
Q\frac{\partial P}{\partial u_n}+(SQ)\frac{\partial P}{\partial
u_{n+1}}+\cdots+(S^{k-1}Q)\frac{\partial P}{\partial u_{n+k-1}}=0.
\end{equation}
\section{Examples}
In this section, we calculate the conservation laws of some
difference equations admitting Lie point symmetries by using our
method. Assume that all singularity conditions are taken into
consideration.

\begin{exm}[\cite{Hydon:2012aa}]
The difference equation
\begin{equation}
u_{n+2}=\frac{u_{n+1}^2}{u_n}
\end{equation}
possessing a symmetry group with characteristics
\begin{equation}
Q_1=u_n,~~Q_2=nu_n,~~Q_3=u_n\ln |u_n|.
\end{equation}
Assume the first integral is $P(n,u_n,u_{n+1})$. The equality
(\ref{pde1}) gives the following linear partial differential
equations
\begin{equation}
Q_j\frac{\partial P}{\partial u_n}+(SQ_j)\frac{\partial P}{\partial
u_{n+1}}=0,~~j=1,2,3.
\end{equation}
When $Q_1$ is substituted, the equation above can be solved that
\begin{equation}
P=P(n,u_{n+1}/u_n).
\end{equation}
Substitute it into the determining condition (\ref{dcffi}) and get
that
\begin{equation}
\begin{aligned}
0=(S-\operatorname{id})P&=P(n+1,u_{n+2}/u_{n+1})-P(n,u_{n+1}/u_n)\\
&=P(n+1,u_{n+1}/u_n)-P(n,u_{n+1}/u_n),
\end{aligned}
\end{equation}
which implies that $P=P(u_{n+1}/u_n)$ is a first integral for any
smooth function $P$. So $u_{n+1}/u_n$ itself is a first integral as
well.

Similarly, when $Q_2$ is used, the partial differential equation can
be solved as
\begin{equation}
P=P(n,u_{n+1}^n/u_n^{n+1}).
\end{equation}
The determining condition amounts to that
$P=P(u_{n+1}^n/u_n^{n+1})$, implying that $u_{n+1}^n/u_n^{n+1}$ is
another first integral and it is functionally independent of
$u_{n+1}/u_n$.

The third characteristic does not provide any more first integrals.
By setting
\begin{equation}
u_{n+1}/u_n=c_1,~~u_{n+1}^n/u_n^{n+1}=1/c_2,
\end{equation}
with $c_1$ and $c_2$ constants, we find the general solution of the
original difference equation (by eliminating $u_{n+1}$)
\begin{equation}
u_n=c_2c_1^n.
\end{equation}

Nevertheless, by assuming $u_n>0$ and setting $s_n=\ln u_n$, this
equation admits a Lagrangian formulation. The Lagrangian is given by
$L_n=s_ns_{n+1}-s_n^2$. It is easy to see that the first two groups
are groups of divergence variational symmetries while the last one
is not by checking (taking Remark \ref{remlc} into consideration)
\begin{equation}
\bold{v}_1\left(L^{(\infty)}\right)=(S-\operatorname{id})s_n,~~
\bold{v}_2\left(L^{(\infty)}\right)=(S-\operatorname{id})((n-1)s_n).
\end{equation}
Here $\bold{v}_1=\partial_{s_n}$ and $\bold{v}_2=n\partial_{s_n}$.
\end{exm}

\begin{exm}[\cite{Hydon:2012aa}]
Consider a third-order difference equation
\begin{equation}
u_{n+3}=u_{n+1}u_{n+2}/u_n.
\end{equation}
It admits a four-dimensional group of Lie point symmetries with
characteristics
\begin{equation}
Q_1=u_n,~~Q_2=(-1)^nu_n,~~Q_3=nu_n,~~Q_4=u_n\ln|u_n|.
\end{equation}
Using $Q_1$, we can get
\begin{equation}
P=P\left(n,\frac{u_{n+2}}{u_{n+1}},\frac{u_{n+1}}{u_n}\right),
\end{equation}
and it satisfies
\begin{equation}
P\left(n+1,\frac{u_{n+1}}{u_n},\frac{u_{n+2}}{u_{n+1}}\right)
=P\left(n,\frac{u_{n+2}}{u_{n+1}},\frac{u_{n+1}}{u_n}\right).
\end{equation}
This yields that $P$ is independent of $n$ and satisfies
\begin{equation}
P(x,y)=P(y,x),~~\text{with}~~x=\frac{u_{n+1}}{u_n}~~\text{and}~~y=\frac{u_{n+2}}{u_{n+1}}.
\end{equation}
Obviously functions of the forms $P(u_{n+2}/u_n)$ and
$P(u_{n+2}/u_{n+1}+u_{n+1}/u_n)$ are solutions and we can get the
corresponding first integrals as follows:
\begin{equation}
P=\frac{u_{n+2}}{u_n},~~P=\frac{u_{n+2}}{u_{n+1}}+\frac{u_{n+1}}{u_n}.
\end{equation}

From the other symmetry characteristics, we cannot find any more
first integrals, which are functionally independent from the ones
obtained from $Q_1$.

Since the order of this equation is odd, there does not exist a
Lagrangian formulation.
\end{exm}

\begin{exm}
Consider the following equation
\begin{equation}
u_{1,1}=\omega(u_{0,0},u_{0,1},u_{1,0})=u_{0,0}\left(1+\frac{u_{0,1}}{u_{1,0}}\right);
\end{equation}
here $u_{0,0}=u(m,n)$ is the value of the dependent variable at the
point $(m,n)\in\mathbb{Z}^2$. Let the shifts of the dependent
variable be $u_{i,j}=S_{(i,j)}u_{0,0}$, or alternatively
$u_{i,j}=S_m^iS_n^ju_{0,0}$ with $S_m$, $S_n$ the unite forward
shifts along the $m$ and $n$ directions. It admits Lie point
symmetries with characteristics
\begin{equation}
Q_1=u_{0,0},~~Q_2=(-1)^{m+n}u_{0,0}.
\end{equation}
Assume a two-tuple $P=(F,G)$ with components
\begin{equation}
F=F(m,n,u_{0,0},u_{0,1})~~\text{and}~~G=G(m,n,u_{0,0},u_{1,0}),
\end{equation}
satisfying
\begin{equation}\label{fg111}
(S_m-\operatorname{id})F+(S_n-\operatorname{id})G=0
\end{equation}
on solutions of the given equation. From (\ref{clss}), we have that
for $Q_1$,
\begin{equation}
F=F(m,n,u_{0,0}/u_{0,1}),~~G=G(m,n,u_{0,0}/u_{1,0}).
\end{equation}
By applying the following differential operators to the divergence
expression (\ref{fg111}) (see Rasin $\&$ Hydon \cite{Rasin:2007aa}),
\begin{equation}\label{l1l20}
\mathcal{D}_1=\frac{\partial}{\partial
u_{0,1}}-\frac{\omega_{u_{0,1}}}{\omega_{u_{0,0}}}\frac{\partial}{\partial
u_{0,0}},~~\mathcal{D}_2=\frac{\partial}{\partial
u_{1,0}}-\frac{\omega_{u_{1,0}}}{\omega_{u_{0,0}}}\frac{\partial}{\partial
u_{0,0}},
\end{equation}
we get that
\begin{equation}\label{l1l2fg}
\mathcal{D}_1\mathcal{D}_2(F+G)=0.
\end{equation} Here $\omega_{u_{i,j}}$ denotes $\frac{\partial \omega}{\partial
u_{i,j}}$. By setting $x=u_{0,0}/u_{0,1}$ and $y=u_{0,0}/u_{1,0}$,
the equation (\ref{l1l2fg}) becomes
\begin{equation}\label{eqfg1}
(2xy+x^2)F''(x)+2yF'(x)-y^2G''(y)-2yG'(y)=0.
\end{equation}
Here the discrete variables are omitted and the derivatives are with
respect to the continuous arguments $x$ and $y$. Differentiating the
last equation with respect to $x$ yields a polynomial with respect
to $y$; coefficients of each power give us the unique equation for
$F$,
\begin{equation}
xF'''(x)+2F''(x)=0.
\end{equation}
This can be easily solved and its general solution
is\footnote{Constant solution is omitted, since it always
contributes to a trivial conservation law, which is not of our
interests.}
\begin{equation}
F(x)=A_1(m,n)\ln x+A_2(m,n)x.
\end{equation}
Substitute this back into (\ref{eqfg1}), and a differential equation
for $G$ can be obtained,
\begin{equation}
y^2G''(y)+2yG'(y)-2A_2(m,n)y+A_1(m,n)=0.
\end{equation}
Its general solution is
\begin{equation}
G(y)=\frac{B(m,n)}{y}-A_1(m,n)\ln y+A_2(m,n)y.
\end{equation}
The result we achieved above is substituted into the divergence
expression (\ref{fg111}) and then the integration constants $A_1$,
$A_2$ and $B$ are determined by solving some simple difference
equations. Finally, we get the following two conservation laws,
\begin{equation}
F=\frac{1+(-1)^{m+n}}{2}\frac{u_{0,0}}{u_{0,1}},~~G=\frac{1+(-1)^{m+n}}{2}\frac{u_{0,0}}{u_{1,0}}+\frac{1-(-1)^{m+n}}{2}
\frac{u_{1,0}}{u_{0,0}}
\end{equation}
and
\begin{equation}
F=\frac{1-(-1)^{m+n}}{2}\frac{u_{0,0}}{u_{0,1}},~~G=\frac{1+(-1)^{m+n}}{2}\frac{u_{0,0}}{u_{1,0}}+\frac{1-(-1)^{m+n}}{2}
\frac{u_{1,0}}{u_{0,0}}.
\end{equation}

The procedure can be similarly applied by using the characteristic
$Q_2$, however, unfortunately we can not obtain any more independent
conservation laws.
\end{exm}

\begin{exm}
Let $m,n$ be the independent variables and $u,v$ be the dependent
ones. Consider a Lagrangian
\begin{equation}
L_n=\frac{u_{1,0}}{v_{0,0}}+\frac{v_{0,1}}{u_{0,0}},
\end{equation}
which admits a group of variational symmetries with infinitesimal
generator
\begin{equation}
\bold{v}=u_{0,0}\partial_{u_{0,0}}+v_{0,0}\partial_{v_{0,0}}.
\end{equation}
This can be easily checked, $\bold{pr}^{(\infty)}\bold{v}(L_n)=0$.
The associated Euler-Lagrange equations are given by the following
partial difference equations
\begin{equation}\label{wuv}
u_{1,1}=\omega^u(u_{0,0},v_{0,1}):=\frac{v_{0,1}^2}{u_{0,0}},~~
v_{1,1}=\omega^v(v_{0,0},u_{1,0}):=\frac{u_{1,0}^2}{v_{0,0}}.
\end{equation}
The simplest conservation laws we may expect are $P=(F,G)$ with
\begin{equation}
F(m,n,u_{0,0},v_{0,0},u_{0,1},v_{0,1})~~\text{and}~~G(m,n,u_{0,0},v_{0,0},u_{1,0},v_{1,0}).
\end{equation}
 By invoking
(\ref{clss}), the components $F$ and $G$ become
\begin{equation}
F\left(m,n,\frac{u_{0,0}}{v_{0,0}},\frac{u_{0,0}}{u_{0,1}},\frac{u_{0,0}}{v_{0,1}}\right),~~
G\left(m,n,\frac{u_{0,0}}{v_{0,0}},\frac{u_{0,0}}{u_{1,0}},\frac{u_{0,0}}{v_{1,0}}\right).
\end{equation}
Obviously $F$ and $G$ contribute to a conservation law if and only
if
\begin{equation}\label{clofpde}
(S_m-\operatorname{id})F+(S_n-\operatorname{id})G=0
\end{equation}
on solutions of the system (\ref{wuv}). By applying the linear
differential operators
\begin{equation}
\mathcal{D}_1=\frac{\partial}{\partial
u_{1,0}}-\frac{\omega^v_{u_{1,0}}}{\omega^v_{v_{0,0}}}\frac{\partial}{\partial
v_{0,0}},~~\mathcal{D}_2=\frac{\partial}{\partial
v_{0,1}}-\frac{\omega^u_{v_{0,1}}}{\omega^u_{u_{0,0}}}\frac{\partial}{\partial
u_{0,0}},
\end{equation}
to (\ref{clofpde}) to eliminate $S_mF$ and $S_nG$, we get
$\mathcal{D}_1\mathcal{D}_2(F+G)=0$ on solutions.
Similarly to the previous examples, we get a nontrivial conservation
law
\begin{equation}
\begin{aligned}
F=\ln\left(\frac{u_{0,0}}{v_{0,0}}\right)+\ln\left(\frac{u_{0,0}}{u_{0,1}}\right),~~
G=-\ln\left(\frac{u_{0,0}}{v_{0,0}}\right)+\ln\left(\frac{v_{0,0}}{v_{1,0}}\right).
\end{aligned}
\end{equation}

The Euler-Lagrange equations possess another group of symmetries,
whose infinitesimal generator is
\begin{equation}
\bold{v}=2nu_{0,0}\partial_{u_{0,0}}+(2n-1)v_{0,0}\partial_{v_{0,0}},
\end{equation}
which obviously does not generate variational symmetries for the
Lagrangian. By the same approach, this provides another conservation
law,
\begin{equation}
F=\frac{(v_{0,1})^{2n}}{(u_{0,0})^{2n+1}},~~G=\frac{(u_{1,0})^{2n-1}}{(v_{0,0})^{2n}}.
\end{equation}
\end{exm}

\begin{exm}
We consider integrable partial difference equations that belong to
the Adler-Bobenko-Suris (ABS) classification \cite{Adler:2003bs}.
The authors obtained their symmetries and conservation laws
respectively in \cite{Rasin:2007sy} and \cite{Rasin:2007aa}. We
simplify their procedure for getting conservation laws using the
obtained Lie point symmetries.

The general form of ABS equations on the quad-graph is
\begin{equation}
\omega(m,n,u_{0,0},u_{1,0},u_{0,1},u_{1,1},\alpha_k,\beta_l)=0.
\end{equation}
The functions $\alpha_m=\alpha(m)$ and $\beta_n=\beta(n)$ play the
roles of edge parameters, which will be chosen as constants in our
case. The equations from the ABS classification are
\begin{equation*}
\begin{aligned}
\bold{A1}:&~~\alpha(u_{0,0}+u_{0,1})(u_{1,0}+u_{1,1})-\beta(u_{0,0}+u_{1,0})(u_{0,1}+u_{1,1})-\delta^2\alpha\beta(\alpha-\beta)=0,\\
\bold{A2}:&~~(\beta^2-\alpha^2)(u_{0,0}u_{1,0}u_{0,1}u_{1,1}+1)+\beta(\alpha^2-1)(u_{0,0}u_{0,1}+u_{1,0}u_{1,1})\\
&~~~~-\alpha(\beta^2-1)(u_{0,0}u_{1,0}+u_{0,1}u_{1,1})=0,\\
\bold{H1}:&~~(u_{0,0}-u_{1,1})(u_{1,0}-u_{0,1})+\beta-\alpha=0,\\
\bold{H2}:&~~(u_{0,0}-u_{1,1})(u_{1,0}-u_{0,1})+(\beta-\alpha)(u_{0,0}+u_{0,1}+u_{1,0}+u_{1,1})+\beta^2-\alpha^2=0,\\
\bold{H3}:&~~\alpha(u_{0,0}u_{1,0}+u_{0,1}u_{1,1})-\beta(u_{0,0}u_{0,1}+u_{1,0}u_{1,1})+\delta^2(\alpha^2-\beta^2)=0,\\
\bold{Q1}:&~~\alpha(u_{0,0}-u_{0,1})(u_{1,0}-u_{1,1})-\beta(u_{0,0}-u_{1,0})(u_{0,1}-u_{1,1})+\delta^2\alpha\beta(\alpha-\beta)=0,\\
\bold{Q2}:&~~\alpha(u_{0,0}-u_{0,1})(u_{1,0}-u_{1,1})-\beta(u_{0,0}-u_{1,0})(u_{0,1}-u_{1,1})\\
&~~~~+\alpha\beta(\alpha-\beta)(u_{0,0}+u_{1,0}+u_{0,1}+u_{1,1})-\alpha\beta(\alpha-\beta)(\alpha^2-\alpha\beta+\beta^2)=0,\\
\bold{Q3}:&~~(\beta^2-\alpha^2)(u_{0,0}u_{1,1}+u_{1,0}u_{0,1})+\beta(\alpha^2-1)(u_{0,0}u_{1,0}+u_{0,1}u_{1,1})\\
&~~~~-\alpha(\beta^2-1)(u_{0,0}u_{0,1}+u_{1,0}u_{1,1})-\delta^2(\alpha^2-\beta^2)(\alpha^2-1)(\beta^2-1)/(4\alpha\beta)=0,\\
\bold{Q4}:&~~\operatorname{sn}(\alpha)(u_{0,0}u_{1,0}+u_{0,1}u_{1,1})-\operatorname{sn}(\beta)(u_{0,0}u_{0,1}+u_{1,0}u_{1,1})-\operatorname{sn}(\alpha-\beta)
(u_{0,0}u_{1,1}+u_{0,1}u_{1,0})\\
&~~~~+\operatorname{sn}(\alpha-\beta)\operatorname{sn}(\alpha)\operatorname{sn}(\beta)(1+K^2u_{0,0}u_{0,1}u_{1,0}u_{1,1})=0.
\end{aligned}
\end{equation*}
Here $\operatorname{sn}(\alpha)=\operatorname{sn}(\alpha;K)$ is a
Jacobi elliptic function with modulus $K$. We write a three-point
conservation law  as
$(F(m,n,u_{0,0},u_{0,1},\alpha,\beta),G(m,n,u_{0,0},u_{1,0},\alpha,\beta))$
such that \begin{equation}\label{pdeex1}
(S_m-\operatorname{id})F+(S_n-\operatorname{id})G=0
\end{equation}
 on solutions of an equation.

Take the equation $\bold{A1}_{\delta=0}$ for example. It possesses
Lie point symmetries, whose characteristics are given by
\cite{Rasin:2007sy}
\begin{equation}
Q_1=(-1)^{m+n},~~Q_2=u_{0,0},~~Q_3=(-1)^{m+n}(u_{0,0})^2.
\end{equation}
It is implied by (\ref{clss}) that conservation laws related with
them are respectively of the forms
\begin{equation}
\begin{aligned}
&F_1(m,n,u_{0,0}+u_{0,1},\alpha,\beta),~~G_1(m,n,u_{0,0}+u_{1,0},\alpha,\beta),\\
&F_2(m,n,u_{0,1}/u_{0,0},\alpha,\beta),~~G_2(m,n,u_{1,0}/u_{0,0},\alpha,\beta),\\
&F_3(m,n,1/u_{0,0}+1/u_{0,1},\alpha,\beta),~~G_3(m,n,1/u_{0,0}+1/u_{1,0},\alpha,\beta).
\end{aligned}
\end{equation}
By assuming that all the singularity conditions are satisfied,
rewrite the equation,
\begin{equation}
\begin{aligned}
u_{1,1}&=\omega(m,n,u_{0,0},u_{0,1},u_{1,0},\alpha,\beta)\\
&=\frac{\beta(u_{0,0}+u_{1,0})u_{0,1}-\alpha(u_{0,0}+u_{0,1})u_{1,0}}
{\alpha(u_{0,0}+u_{0,1})-\beta(u_{0,0}+u_{1,0})}.
\end{aligned}
\end{equation}
Applying similar differential operators $\mathcal{D}_1$ and
$\mathcal{D}_2$ as given by (\ref{l1l20}) to the divergence
expression (\ref{pdeex1}), we get
\begin{equation}
\mathcal{D}_1\mathcal{D}_2(F_k+G_k)=0,~~k=1,2,3.
\end{equation}
When $k=1$, this gives
two conservation laws
\begin{equation}
\begin{aligned}
F_1=-(-1)^{m+n}\beta(u_{0,0}+u_{0,1})^{-1},~~G_1=(-1)^{m+n}\alpha(u_{0,0}+u_{1,0})^{-1}
\end{aligned}
\end{equation}
and
\begin{equation}
\begin{aligned}
F_1=-(-1)^{m+n}(2\ln(u_{0,0}+u_{0,1})-\ln\beta),~~G_1=(-1)^{m+n}(2\ln(u_{0,0}+u_{1,0})-\ln\alpha).
\end{aligned}
\end{equation}
We list all the characteristics associated with Lie point symmetries
and the forms of their related three-point conservation laws (the
arguments $m,n,\alpha,\beta$ are omitted). A similar procedure as
for $\bold{A1}_{\delta=0}$ (as well as for the other previous
examples) will give us the conservation laws explicitly for all the
equations from the ABS classification as shown in
\cite{Rasin:2007aa}. This method can also be applied to find
 conservation laws depending on more than three points.
\begin{table}[ht]
\centering 
\begin{tabular}{l l l l} 
\textbf{Equations} & \textbf{Characteristics} & \textbf{Potential conservation Laws} \\ [0.5ex] 
\hline 
$\bold{A1}_{\delta=0}$ & $Q_1=(-1)^{m+n}$ & $F_1(u_{0,0}+u_{0,1})$, $G_1(u_{0,0}+u_{1,0})$  \\
                       & $Q_2=u_{0,0}$ & $F_2(u_{0,1}/u_{0,0})$, $G_2(u_{1,0}/u_{0,0})$  \\
                       & $Q_3=(-1)^{m+n}(u_{0,0})^2$ & $F_3(1/u_{0,0}+1/u_{0,1})$, $G_3(1/u_{0,0}+1/u_{1,0})$  \\
$\bold{A1}_{\delta=1}$ & $Q_1=(-1)^{m+n}$ & $F_1(u_{0,0}+u_{0,1})$, $G_1(u_{0,0}+u_{1,0})$  \\
                       & $Q_2^u=u_{0,0},~Q^{\alpha}_2=\alpha,~Q^{\beta}_2=\beta$ & $F_2(u_{0,1}/\alpha,u_{0,0}/\alpha,\beta/\alpha)$, $G_2(u_{1,0}/\alpha,u_{0,0}/\alpha,\beta/\alpha)$  \\
$\bold{A2}$            & $Q=(-1)^{m+n}u_{0,0}$ & $F(u_{0,0}u_{0,1})$, $G(u_{0,0}u_{1,0})$ \\
[1ex] 
\hline 
\end{tabular}
\label{table:nonlin} 
\end{table}

\begin{table}[ht]
\centering 
\begin{tabular}{l l l l} 
\textbf{Equations} & \textbf{Characteristics} & \textbf{Potential conservation Laws} \\ [0.5ex] 
\hline 
$\bold{H1}$            & $Q_1=1$ & $F_1(u_{0,1}-u_{0,0})$, $G_1(u_{1,0}-u_{0,0})$\\
                       & $Q_2=(-1)^{m+n}$ & $F_2(u_{0,0}+u_{0,1})$, $G_2(u_{0,0}+u_{1,0})$\\
                       & $Q_3^u=u_{0,0},~Q_3^{\alpha}=2\alpha,~Q_3^{\beta}=2\beta$ & $F_3((u_{0,1})^2/\alpha,(u_{0,0})^2/\alpha,\beta/\alpha)$,\\ &&$G_3((u_{1,0})^2/\alpha,(u_{0,0})^2/\alpha,\beta/\alpha)$\\
                       & $Q_4=(-1)^{m+n}u_{0,0}$ & $F_4(u_{0,0}u_{0,1})$, $G_4(u_{0,0}u_{1,0})$\\
$\bold{H2}$            & $Q_1^u=1,~Q_1^{\alpha}=-2,~Q_1^{\beta}=-2$ & $F_1(2u_{0,1}-\alpha,2u_{0,0}-\alpha,\beta-\alpha)$, \\
 &&$G_1(2u_{1,0}-\alpha,2u_{0,0}-\alpha,\beta-\alpha)$\\
                       & $Q_2=(-1)^{m+n}$ & $F_2(u_{0,0}+u_{0,1})$, $G_2(u_{0,0}+u_{1,0})$\\
                       & $Q_3^u=u_{0,0},~Q_3^{\alpha}=\alpha,~Q_3^{\beta}=\beta$ & $F_3(u_{0,1}/\alpha,u_{0,0}/\alpha,\beta/\alpha)$, $G_3(u_{1,0}/\alpha,u_{0,0}/\alpha,\beta/\alpha)$\\
$\bold{H3}_{\delta=0}$ & $Q_1=u_{0,0}$ & $F_1(u_{0,1}/u_{0,0})$, $G_1(u_{1,0}/u_{0,0})$\\
                       & $Q_2=(-1)^{m+n}u_{0,0}$ & $F_2(u_{0,0}u_{0,1})$, $G_2(u_{0,0}u_{1,0})$\\
$\bold{H3}_{\delta=1}$ & $Q_1^u=u_{0,0},~Q_1^{\alpha}=2\alpha,~Q_1^{\beta}=2\beta$ & $F_1((u_{0,1})^2/\alpha,(u_{0,0})^2/\alpha,\beta/\alpha)$,\\ &&$G_1((u_{1,0})^2/\alpha,(u_{0,0})^2/\alpha,\beta/\alpha)$\\
                       & $Q_2=(-1)^{m+n}u_{0,0}$ & $F_2(u_{0,0}u_{0,1})$, $G_2(u_{0,0}u_{1,0})$\\
$\bold{Q1}_{\delta=0}$ & $Q_1=1$ & $F_1(u_{0,1}-u_{0,0})$, $G_1(u_{1,0}-u_{0,0})$\\
                       & $Q_2=u_{0,0}$ & $F_2(u_{0,1}/u_{0,0})$, $G_2(u_{1,0}/u_{0,0})$\\
                       & $Q_3=(u_{0,0})^2$ & $F_3(1/u_{0,0}-1/u_{0,1})$, $G_3(1/u_{0,0}-1/u_{1,0})$\\
$\bold{Q1}_{\delta=1}$ & $Q_1=1$ & $F_1(u_{0,1}-u_{0,0})$, $G_1(u_{1,0}-u_{0,0})$\\
                       & $Q_2^u=u_{0,0},~Q_2^{\alpha}=\alpha,~Q_2^{\beta}=\beta$ & $F_2(u_{0,1}/\alpha,u_{0,0}/\alpha,\beta/\alpha)$, $G_2(u_{1,0}/\alpha,u_{0,0}/\alpha,\beta/\alpha)$\\
$\bold{Q2}$            & $Q^u=2u_{0,0},~Q^{\alpha}=\alpha,~Q^{\beta}=\beta$ & $F(u_{0,1}/\alpha^2,u_{0,0}/\alpha^2,\beta/\alpha)$, $G(u_{1,0}/\alpha^2,u_{0,0}/\alpha^2,\beta/\alpha)$\\
$\bold{Q3_{\delta=0}}$ & $Q=u_{0,0}$ & $F(u_{0,1}/u_{0,0})$, $G(u_{1,0}/u_{0,0})$\\
$\bold{Q4}_{K=\pm1}$   & $Q=1-(u_{0,0})^2$ &
                         $F\left(\frac{(u_{0,0}+1)(u_{0,1}-1)}{(u_{0,0}-1)(u_{0,1}+1)}\right)$,
                         $G\left(\frac{(u_{0,0}+1)(u_{1,0}-1)}{(u_{0,0}-1)(u_{1,0}+1)}\right)$\\
[1ex] 
\hline 
\end{tabular}
\label{table:nonlin} 
\end{table}

\end{exm}


 \acknowledgements This work was
supported by the Department of Mathematics at University of Surrey,
the China Scholarship Council and JST-CREST. The author would like to thank
Professor Peter Hydon for
valuable comments on an earlier version.

\bibliographystyle{latexeu}

\bibliography{references}

\end{document}